# Revealing the online network between University and Industry: The case of Turkey

Enrique Orduna-Malea[1*] and Selenay Aytac[2]

[1] *EC3 Research Group, Polytechnic University of Valencia. Camino de Vera s/n, Valencia 46022, Spain*
[2] Long Island University, 720 Northern Blvd., Brookville, NY 11548, USA; E-Mail: selenay.aytac@liu.edu

\* enorma@upv.es

**Abstract** The present paper attempts to explore the relationship between the Turkish academic and industry systems by mapping the relationships under web indicators. We used the top 100 Turkish universities and the top 10 Turkish companies in 10 industrial sectors in order to observe the performance of web impact indicators. Total page count metric is obtained through Google Turkey and the pure link metrics have been gathered from Open Site Explorer. The indicators obtained both for web presence and web visibility indicated that there are significant differences between the group of academic institutions and those related to companies within the web space of Turkey. However, this current study is exploratory and should be replicated with a larger sample of both Turkish universities and companies in each sector. Likewise, a longitudinal study rather than sectional would eliminate or smooth fluctuations of web data (especially URL mentions) as a more adequate understanding of the relations between Turkish institutions, and their web impact, is reached.

## 1. Introduction

Following teaching and research, transfer activities are one of the most essential part of the third university mission (Montesinos et al. 2008). Technology transfer between universities and companies are a prominent example, resulting in transfer of knowledge to industry and contributing to the economic development of a country.

Generally, these transfer processes are measured through patent analysis or patentometrics (Meyer 2000). The ability to share the authorship of inventions and to transfer improvements in these processes, methodologies, and services may be intensified by an entrepreneurial university model (Meyer, Siniläinen and Utecht 2003).

Bibliometrics also allows us to partially measure some of these interactions through the analysis of the scientific contributions co-authored by both university researchers and private company workers. Two recent examples are the SCImago Institutions Ranking[1] and the UIRC (University - Industry Research Connections).[2] SCImago provides bibliometric data about the research, innovation, and web visibility of the private sector (taken from Scopus), and the UIRC (University - Industry Research Connections) allows the mapping of relationships between universities and business, from data of the world's 750 largest research universities in the Leiden Ranking 2014, providing indicators of intensity, local industrial partners, domestic industrial partners, and foreign industrial partners.

When we add a third node to these two entities (universities and companies), corresponding to the Administration, we pass to the Triple-Helix, which is a concept popularized by Etzkowitz and Leydesdorff (1996). Although measurement of synergies raises some methodological concerns (Leydesdorff and Park 2014), Triple-Helix allowed us to measure the intensity of the inter-relationships between these three social engines, under the assumption that, in advanced societies, these interactions are of great intensity.

The emergence and subsequent development of hyperspace allowed the possibility of explicit relationships (through hyperlinks, textual mentions or keyword queries) both formal and informal, which cannot be observable neither in the patent nor bibliometric environments. For instance, the link analysis (Thelwall 2004) began to be applied in academic environments, such as universities (Aguillo, Granadino, Ortega and Prieto 2006), academic journals (Vaughan and Thelwall 2003) and, more recently, the social web arena (Priem and Hemminger 2010; Thelwall 2014). On the other hand, the application of web indicators on the performance of companies through cybermetric techniques (to which link analysis belongs to) is relatively new when compared with other applied areas of this discipline.

During the last decade a limited but growing research field which aims to study commercial companies has emerged (Vaughan 2004; Romero-Frias 2011). Some of these studies focused on expanding precise triple-helix analyses by adapting and applying them in a web environment, using techniques taken from cybermetrics. In this vein, we can highlight the analysis of the automotive industry in the United Kingdom (Stuart and Thelwall 2006) or the study performed by Khan and Park (2011), who detected the influence of sources and policies in the intensity of relationships. Furthermore, Garcia-Santiago and Moya-Anegón (2009) utilized co-outlink techniques in order to study if they could visualize the triple-helix structure in different sectors such as business associations, banks, ministries, chambers of commerce, foundations, port authorities, public health centers, laboratories, mass media, or universities. These studies suggested that the Web is "reconfirmed as the faithful mirror image of official communications among organisms that form part of the triple helix", though only those organisms that are fully institutionalized actually facilitate stable processes of cooperation. Minguillo and Thelwall (2012) also analysed the network structure of science parks as a node between universities and industry, and have found that cybermetric techniques assist in discovering patterns that help gain deeper insights into how organisations engage on the Web, and how link analysis may provide evidence about their offline relationships.

Nonetheless, the usage of the web by universities and companies in creating, sharing, disseminating and consuming information online is completely different due to the diverse organizational cultures of these institutions (Vaughan and Wu 2004). Therefore, web indicators should be properly contextualised when applying them in each of these environments. Moreover, possible correlations may not be revealed between existing indicators in an academic environment, an industrial environment, or vice versa. For that reason, these aspects should be considered when measuring the interrelationships between universities and industry. Web relationships heavily depend on linguistic, cultural, and proximity factors (Vaughan 2006) and in the specific case of companies, they might also depend on industrial sectors (Vaughan & Romero-Frias 2012).

There are still few studies that have attempted to compare the correlations between various web indicators depending on the academic or industrial set of institutions in specific places. It should be mentioned the valuable work carried out by Vaughan and Yang (2012), who compared various web indicators in both environments (academic and industrial) in two different countries (US and China), that the correlations are always positive and finding significant though slightly higher in the academic environment.

This is of particular importance in countries with large and complex academic and industrial systems, such as in Turkey. Some scarce studies of a cybermetric nature have been performed, but do not offer correlations between the web indicators used, apart from the inclusion of its university system in the Ranking Web of World Universities[3] or the recent analysis of the online reputation index of Turkish Universities (Arslan and Seker 2014) and companies (Cankir, Arslan and Seker 2015), which analyses the web reputation through various parameters (Google, Facebook, Twitter, etc.) but do not offer correlations between the web indicators used.

The main purpose of this study research is to describe and compare the academic and industry Turkish systems on the web, and map the relationships between them under the web indicators. Consequently, this study will seek an answer to the following research questions:
  RQ1: Are there significant differences between the academic and industrial Turkish system in the performance of web impact indicators (web presence and web visibility)?
  RQ2: Are there significant differences in the correlation of web impact indicators for the academic and industrial Turkish systems?
  RQ3: Is there a significant relationship between the academic and industrial Turkish systems on the web; and how does the industrial sector influence them?

## 2. Methodology

In order to answer our research questions, first we proceeded to obtain the sample object of study both for Turkish universities and companies. Secondly the data sources were selected and web indicators were obtained, and finally a statistical analysis was conducted.

In the case of universities we decided to select the Top 100 Turkish universities according to the latest available edition of the Ranking Web of Universities (July 2014 Edition). Despite the existence of other university rankings with extensive coverage in Turkey, such as the University Ranking by Academic Performance (URAP),[4] the selection of the Ranking Web was motivated by its being a classification based on web indicators, ranking Turkish universities according to their performance on the web, precisely the dimension to be evaluated in this study. On the other hand, the scientific literature has demonstrated the positive correlation between web indicators and the academic performance of universities (Thelwall and Harries 2003; Aguillo, Granada, Ortega, and Prieto 2006), reinforcing the choice of this ranking as a basis for obtaining samples for quantitative studies.

We used a sample of 100 companies from the top 10 valuated companies in each of 10 major industrial sectors, from the latest available edition of "Turkey's Top 500 Industrial Enterprises", where companies are ranked according to the "Production-based sales" indicator prepared by the Istanbul Chamber of Industry (Bahcivan 2013). The choice of this source was motivated by the need to identify the most powerful national companies in different industries and, which are, more likely to participate in collaborative activities within the university sector. Additionally, the scientific literature has demonstrated the existence of a moderate positive correlation between certain financial variables and web indicators (Vaughan 2004), especially in groups of companies belonging to homogeneous sectors. Therefore, the selection of companies a) according to sector; and b) based on some financial variable, is considered timely in order to obtain a sample of companies with great performance on the Web.

The URL for each university and company was finally located. Two companies were identified with more than one URL: Philsa Philip Morris Sabancı Sigara ve Tütüncülük San. ve Tic. A.Ş (<philsa.com.tr> and <pmkariyer.com>), and Arçelik A.Ş (<arcelikas.com> and <arcelik.com.tr>). For web impact measures, only the URL with better page count was used; content duplicated and self-mentions between each URL make useless the combination of these URLs for web performance purposes.

The final samples of universities and companies are available in Annexes I and II respectively, in the supplementary material 1.[5]

## 2.1. Web impact measures

After obtaining the sample, the web indicators and sources necessary to answer research questions 1 and 2 were determined and applied. These web metrics, their scope and source are showed in Table 1. When corresponding, the necessary query to obtain the indicator is offered as well.

**Table 1. Web sources and metrics**

| METRIC | SCOPE | SOURCE | QUERY |
|---|---|---|---|
| Total page count (TPC) | Number of files indexed within a web domain | Google | site:abc.com |
| Academic page count (APC) | Number of files indexed within a web domain | Google Scholar | site:abc.com |
| Citations | Number of citations received by an institution (2000-2013) | Scopus | (AFFILCOUNTRY(turkey) AND PUBYEAR=xxxx |
| External inlinks | Number of times a one URL is linked from Any external website | OSE | Direct |
| Root domains | Number of unique root domains containing at least one link to an specific URL | OSE | Direct |
| Domain authority | Web domain popularity score (0 to 100) | OSE | Direct |
| Global URL mention (GUM) | Number of times a one URL is mentioned outside its specific URL | Google | "abc.com" -site:abc.com Region = All |
| Local URL mention (LUM) | Number of times a one URL is mentioned outside its specific URL | Google | "abc.com" -site:abc.com Region = Turkey |

The total page count metric is obtained through Google Turkey version (<google.com.tr>) by using the "site" command; Google's choice is motivated by being the commercial search engine with the most comprehensive index of the web today. Although the "site" command is not exhaustive,[6] the possible error rate introduced is equally distributed to all web domains, so its effect is minimized if data are taken in a comparative manner.

Since Google does not provide link functionality at present, the "URL mentions" have been calculated as an alternative for link metrics; this indicator has already been successfully tested as an accurate substitute (Thelwall, Sud and Wilkinson 2012; Ortega, Orduna-Malea and Aguillo 2014). In order to get insights about the internationality of links, the URL mentions have been calculated both globally (mentions coming from elsewhere) and locally by selecting the region of Turkey in the advanced settings (mentions coming only from websites hosted in Turkey).

In any case, pure link metrics have been obtained from Open Site Explorer (OSE),[7] although its lower coverage if compared with Google must be taken into account. The following link-related metrics have been gathered:

(1) External inlinks (link-level metric): number of external links received by the web domain analyzed.
(2) Root domains linking (site-level metric): it quantifies the amount of websites that link to one URL instead of quantifying the total amount of external links; and
(3) Domain authority (weighted-level metric); this indicator reflects the reputation of each web domain in a similar fashion as PageRank does, using a more discriminating range (0 to 100).

In the case of universities, academic-related content was additionally gathered for a better characterization of these institutions. The total number of scholarly productivity data was obtained from Scopus. As many of the institutions in the sample were created recently (less than 10 years), citations were restricted from 2000 to 2013 for a better comparison of current academic performance. As a complement, academic page count was retrieved from Google Scholar (with "site" command). As the custom range of Google Scholar does not work accurately at present, no data restriction was performed.

**2.2. Network measures**

An online network between universities and companies is needed to give a proper answer to research question 3. Due to the complexity of this operation, only the Top 25 universities in the Ranking Web of Universities were selected, although all 100 companies were considered. The connection between a university and a company was measured by specific URL mentions. For example, the query <"abc.com" site:xyz.com> retrieves approximately the number of times that the university website <abc.com> is mentioned on the company website <xyz.com>.

Two different companies were identified as belonging to different industrial sectors in the sample, but holding the same general URL: <icdas.com.tr> (İçdaş Elektrik Enerjisi Üretim ve Yatırım; *İçdaş Çelik Enerji Tersane ve Ulaşım Sanayi A.Ş.*) and <zorlu.com.tr> (Zorluteks Tekstil Tic. ve San. A.Ş.; *Vestel Beyaz Eşya San. ve Tic. A.Ş.*). In this case, each URL was assigned only to one company (that was in the better position on the ranking, and marked previously by italics). Another company (Soma Kömür İşletmeleri A.Ş.) is currently out of order so it was also excluded from the network, which was finally composed of 25 universities and 98 companies.

The queries for all possible combinations (15,750) were conducted via Google directly. This procedure was automated to scrape the hit count estimates (HCE) from the first search engine page result (SERP). Google was selected due to its higher coverage and the current inability of Bing API to obtain HCE for web domains over 1,000 hits.

The data were exported into a spreadsheet from which a NET file was manually built, and then exported to Gephi v. 0.8.2, from which we obtained directly both node-level indicators (degree, closeness, betweenness, eigenvector, clustering coefficient), and network-level indicators (average degree, diameter, density, average clustering coefficient, average path length), for each institution (Table 2). Net visualization was generated by the Fruchterman & Reingold force-directed graph drawing algorithm (Fruchterman & Reingold 1991).

Table 2. Metrics for university-company URL mention network

| NODE LEVEL | SCOPE |
| --- | --- |
| **Degree** | Number of edges that are adjacent to one node |
| **Closeness** | How often a node appears on shortest paths between nodes in the network |
| **Betweenness** | The average distance from a given node to all other nodes in the network |
| **Clustering coefficient** | The degree to which nodes of the neighborhood of a node "a" are connected each other |

| NETWORK LEVEL | SCOPE |
| --- | --- |
| **Avg. degree** | The average degree over all of the nodes in the network |
| **Avg. clustering coefficient** | The average clustering coefficient over all of the nodes in the network |
| **Avg. path length** | The average graph-distance between all pairs of nodes |
| **Diameter** | The maximal distance between all pairs of nodes |
| **Graph density** | How close the network is to complete (density equal to 1) |

## 2.3. Statistical analysis

All data (web impact measures and specific URL mentions for the network) were manually retrieved during the last week of December 2014 and then exported to XLstat to perform the following statistical analyses:

- Descriptive statistics: for each set of institutions (universities and companies), the mean, median, standard deviation, kurtosis (Spearman) and skewness (Spearman) of each web indicator were obtained to compare possible differences.
- Correlation analysis: web impact indicators were correlated with university and company sets separately in order to find possible differences in each environment. Since web data presents a skewed distribution (Barabasi and Albert, 1999), Spearman ($\alpha=0.01$) was applied in all calculations.
- Principal Component Analysis (PCA): it was performed in order to complement correlation by finding causes that explain the variability of the indicators applied to the samples of universities and companies (Jolliffe 2002). The Pearson (n) PCA with varimax rotation was applied. All data (with the exception of domain authority) were log-transformed prior to analysis. PCA has been widely used in webometric analysis since it allows identifying a small number of factors that explain most of the variance observed in a greater number of variables. The first factors already explain a major percentage of the variance, while the last factors generally explain very little of the variance so that they can be discarded without the risk of losing much information (Faba-Fernández, Guerrero-Bote & Moya-Anegón 2003).

## 3. Results

### 3.1. Web impact of Turkish universities and companies

The results of descriptive statistical analysis for all universities and companies are offered in Table 3. For reasons of text space, the full details of each institution in each of the indicators are available in Annexes I (universities) and II (companies).

**Table 3. Descriptive statistics for university and company sets**

| WEB METRICS | Median | Mean | Standard deviation | Kurtosis (Pearson) | Skewness (Pearson) |
|---|---|---|---|---|---|
| | | | | | **UNIVERSITY SET** |
| TPC | 89,550.0 | 314,368.3 | 996,643.5 | 37.8 | 6.1 |
| APC | 651.0 | 1,454.1 | 2,719.7 | 22.6 | 4.3 |
| GUM | 117,500.0 | 194,753.0 | 301,489.3 | 49.3 | 6.4 |
| LUM | 62,150.0 | 102,973.4 | 121,317.1 | 7.8 | 2.6 |
| Domain authority | 57.5 | 57.2 | 6.9 | 0.0 | 0.5 |
| External links | 14,342.0 | 33,211.5 | 59,128.6 | 28.5 | 4.8 |
| Root domain | 1,157.0 | 1,632.4 | 1,590.2 | 5.7 | 2.3 |
| | | | | | **COMPANY SET** |
| WEB METRICS | Median | Mean | Standard Deviation | Kurtosis (Pearson) | Skewness (Pearson) |
| TPC | 628.0 | 1,754.9 | 4,008.0 | 23.0 | 4.6 |
| APC | 0.0 | 0.2 | 1.1 | 69.0 | 7.9 |
| GUM | 10,400.0 | 25,504.3 | 36,465.5 | 7.9 | 2.6 |
| LUM | 4,630.0 | 14,865.4 | 29,459.6 | 14.8 | 3.6 |
| Domain authority | 33.0 | 32.9 | 10.6 | 0.4 | 0.2 |
| External links | 1,215.0 | 8,713.9 | 35,674.6 | 56.5 | 7.2 |
| Root domain | 81.0 | 198.2 | 679.6 | 81.7 | 8.9 |

The data relating to page count reveal a significant difference in web presence. While the median value for universities is 89,550 hits, for companies this value is reduced to 628 hits. These results are exemplified by the institutions that have achieved a greater presence.

In the case of universities, these are Suleyman Demirel University Turkey (7.09 million), followed by Istanbul University (6.99 million) and Bilkent University (2.05 million). The 49% of universities have over 100,000 results while only 7% get less than 10,000. However, page count values on companies are orders of magnitude lower. Arçelik A.Ş. is the company that manages a highest page count (27,400), followed by Kaleseramik Çanakkale Kalebodur Seramik Sanayi A.Ş. (22,400) and Mercedes-Benz T.A.Ş. (17,500). It is significant that 21% of companies do not exceed 100 hits.

These page count data clearly condition the web visibility results, reflected in the differences obtained in the number of root domains. While 59% of universities exceed the number of 1,000 root domains from which they receive at least one external link (being the Middle East Technical University, the first with 8,577 root domains), only one company (Feza Gazetecilik A.Ş.) exceeds this threshold (6,701), although it should be clarified that this value corresponds to the general web domain of the group <zaman.com.tr> (domain authority values are only calculated for the top level domains), not to the specific company <ik.zaman .com.tr>, which only gets in fact 4 root domains. The domain authority (which distributes normally) median value is equally significant; 57.5 for universities, and 33 for companies.

Additionally, we have gathered page count data from Google Scholar for both sets of institutions. In the case of universities, Istanbul University is the institution that achieves the highest performance (20,200), followed by Atatürk University (11,200). These are the only universities that exceed the threshold of 10,000 hits. Nevertheless, the contribution of these results over their total page count is low (0.3% and 5.5% respectively). In aggregate data, of the 31,436,830 pages indexed by the 100 universities analysed, only 0.5% are indexed in Google Scholar.

Although the research is not one of the main missions of companies, it was considered appropriate to gather the academic page count in these institutions as well. The only company that highlights is Eti Maden İşletmeleri Genel Müdürlüğü (10 records). Behind this, two companies returned two records each, and eight companies one record; the remaining companies (89%) have not indexed any record in Google Scholar.

Finally, the URL mention provides excessively high results. In the case of universities, the median applied to all 100 universities reached the figure of 117,500. This classification is led by Çankaya University (2.76 million), followed by Mehmet Akif Ersoy University (1.13 million) and Ahi Evran University (567,000), all of them universities located in backward positions in the Top 100 Ranking Web of Universities. For companies, the results are equally high. In five companies the value of 100,000 hits is exceeded, Hürriyet Gazetecilik ve Matbaacılık A.Ş. occupying the first position (205,000), followed by Hayat Kimya Sanayi A.Ş. (187,000), and Çimsa Çimento San. ve Tic. A.Ş. (109,000).

Regional data show very mixed results for their part. In some cases, the results restricted to the region of Turkey constitute elevate percentages of the total URL mentions, such as Suleyman Demirel University Turkey (97.1%) or the company Menderes Tekstil San. ve Tic. A.Ş. (91.4%), and very low percentages in other cases, such as Çankaya University (2.2%) or the company Çayeli İşletmeleri Bakır A.Ş. (0.4%). Globally, the regional results constitute 52.9% of the total number of URL mentions received by all universities; and 58.3% for all companies.

### 3.2. Correlation between web impact metrics for university and company sets

Both page count and web visibility indicators (URL mention, domain authority, external inlinks and root domains), and also citations received, show a high degree of correlation between them (Table 4).

Table 4. Correlation matrix and PCA for Turkish universities (n=100)

|  | TPC | APC | Citations | GUM | Dom. authority | Ext. Links | Root domain |
|---|---|---|---|---|---|---|---|
| *TPC* | 1 | | | | | | |
| *APC* | **0.67 | 1 | | | | | |
| *Citations* | **0.56 | **0.66 | 1 | | | | |
| *GUM* | **0.66 | **0.73 | **0.69 | 1 | | | |
| *Dom. authority* | **0.54 | **0.56 | **0.74 | **0.63 | 1 | | |
| *Ext. links* | **0.56 | **0.53 | **0.58 | **0.62 | **0.70 | 1 | |
| *Root domain* | **0.59 | **0.62 | **0.78 | **0.66 | **0.92 | **0.73 | 1 |

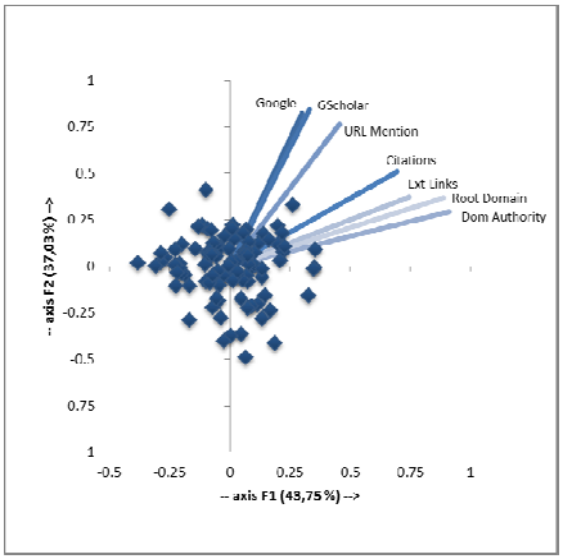
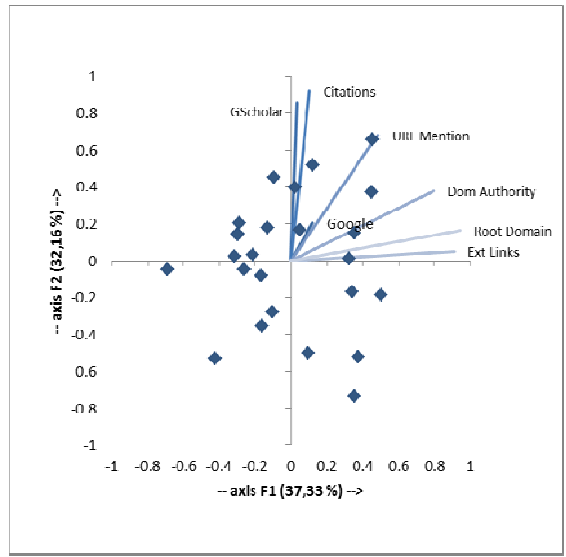

| n=100 universities | n=25 universities |

**\*\*** Significant values (except diagonal) at the level of significance alpha=0.01 (two-tailed test)

The highest correlation is that obtained between the domain authority and doot domain (r = 0.92), although the relation between them is determined by using the root domain in the calculation of authority. Also we should note the high correlation value of citations received with other web indicators, especially with the root domain (r = 0.78).

Although the correlation obtained between total page count and academic page count (Scholar) is high though lower than expected (r = 0.67), the PCA (Table 4) confirms a separation of web presence from web visibility (domain authority, root domain and external inlinks). Moreover, the proximity of the citations received to the web visibility metrics confirms the relationship of these indicators with the impact, broadly understood (impact of research, impact of web content).

The URL mentions deserve special attention; despite achieving significant correlations with visibility indicators such as domain authority (r = 0.63), external links (r = 0.62) and root domains (r = 0.66), they unexpectedly obtain a higher correlation with academic page count in Google Scholar (r = 0.73), fact that causes its closest position in the PCA to the page count metrics than to the web visibility metrics, whilst being essentially an indicator of this nature.

Nonetheless, when only Top 25 universities are considered (Table 4 down; right), the general correlation values between web presence and web impact values down significantly, and conversely, citations get a more central importance, maintaining positive correlations both with Google Scholar's page count (r= 0.68) and URL mention (r = 0.65).

In the case of companies (Table 5), the correlations are also elevated, although several significant differences are observed. On the one hand, we have omitted the indicators related to academic content (Google Scholar, Scopus), since they are practically nil. On the other hand, data about product-based sales have been included.

**Table 5. Correlation matrix and PCA for Turkish companies**

|  | *TPC* | *GUM* | *Dom. authority* | *Ext. links* | *Root domain* | *Sales* |
|---|---|---|---|---|---|---|
| *TPC* | 1 | | | | | |
| *GUM* | **0.50 | 1 | | | | |
| *Dom. authority* | **0.70 | **0.35 | 1 | | | |
| *Ext. links* | **0.66 | **0.36 | **0.85 | 1 | | |
| *Root domain* | **0.72 | **0.37 | **0.96 | **0.82 | 1 | |
| *Sales* | **0.31 | 0.06 | **0.43 | *0.26 | **0.43 | 1 |

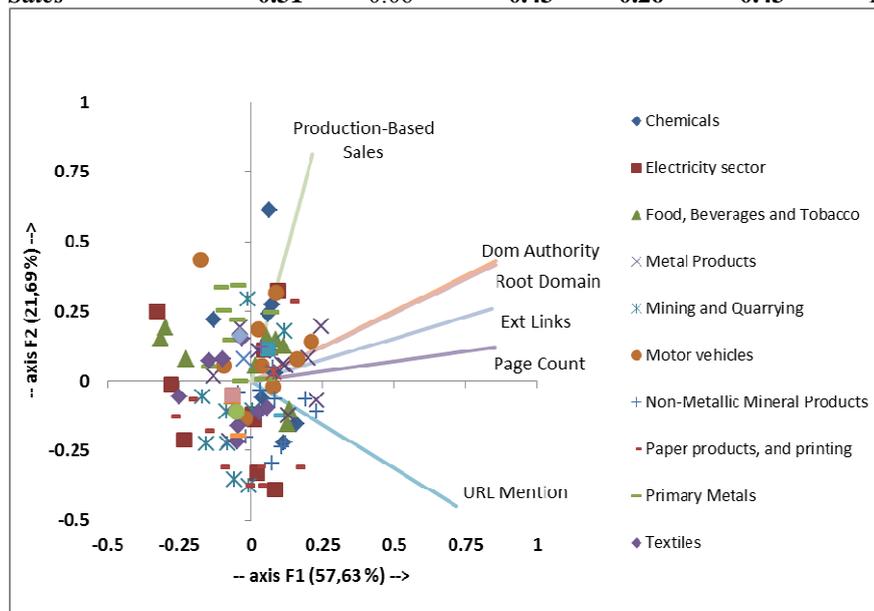

** Significant values (except diagonal) at the level of significance alpha=0.01 (two-tailed test)
 * Significant values (except diagonal) at the level of significance alpha=0.05 (two-tailed test)

In Table 5 we can observe how the page count correlates more significantly with the web visibility indicators, namely domain authority (r = 0.7), external inlinks (r = 0.66) and root domains (r = 0.72). Otherwise, low correlations of production-based sales with all web indicators are identified. As regards the URL mentions, low correlations with the web visibility indicators are detected as well. This effect, previously detected in the case of universities, is accentuated in the case of companies, being the root domain the web visibility indicator that best correlates with the URL mention (r = 0.37).

These results explain the positions shown in the PCA, where web visibility indicators based on links and page count appear closer to each other while URL mentions and production-based sales appear completely separated, reflecting different dimensions of the bodies analysed.

However, if the analysis is performed by industrial activity sectors, the results change significantly. For example, the correlation between page count and root domains is very high for "Non-metallic mineral products" (r = 0.92), "Metal products, machinery and equipment, professional instruments" (r = 0.89) or "Motor vehicles" (r = 0.87), but is very low for "Paper, paper products, and printing" (r = 0.13), "Textiles, wearing apparel, leather and footwear" (r = 0.32) and "Primary metals" (r = 0.34). In the case of URL mentions and root domains, the dependence on the sector is equally remarkable, being elevated for "Metal products, machinery and equipment, professional instruments" (r = 0.74) and "Textiles, wearing apparel, leather and footwear" (r = 0.73), moderate for "Motor vehicles" (0.62), and very low in other sectors. Full details of the industrial sector correlations are available in Annex III of the supplementary material 1.

### 3.3. University-Company interaction network

The network between universities and companies obtained through URL mentions is very compact (graph density = 0.077), with a large network diameter (D = 6) and a moderate average clustering coefficient (0.536). The main network parameters are summarized in Table 6.

**Table 6. Network metrics**

| NET-LEVEL METRIC | VALUE |
|---|---|
| Average degree | 9.425 |
| Diameter | 6 |
| Graph density | 0.077 |
| Average path-lengh | 2.43 |
| Average clustering coefficient | 0.536 |

We can examine the obtained network in Figure 1 (full NET file containing the raw data is provided in the supplementary material 2), where you can observe the high connectivity of universities, located in the center of the graph, with a high web presence (page count proportional to the diameter of each node) and highly connected nodes. By contrast, companies are positioned on the periphery of the graph, surrounding the set of academic nodes; they are less central and with lower page count, node connectivity and cohesion, both among themselves and between them and the university nodes.

In terms of different industries, no specific patterns of centrality except for specific cases (<kale.com.tr>; <Arcelik.com.tr>; <etimaden.gov.tr>; <siemens.com.tr >) are found. If anything a larger web presence and centrality in much of the companies belonging to the "Chemicals industry, petroleum products, rubber and plastics", and for the "Metal products, machinery and equipment, professional instruments" is appreciated.

**Figure 1. University and company interaction network (n=123; algorithm: Fruthterman-Reingold)**

**Legend**: blue: universities; Light Brown: Food, Beverages and Tobacco; Dark Brown: Mining and Quarrying; Green: Chemicals, Petroleum products, Rubber and plastics; Cyan: Metal Products, Machinery and Equipment, Professional instruments; Red: Motor vehicles; Purple: Non-Metallic Mineral Products; Pink: Electricity sector; Grey: Paper, Paper products, and printing; Yellow: Primary Metals; Orange: Textiles, Wearing Apparel, Leather and Footwear.

The node with a higher betweenness of the network corresponds to Çukurova University <cu.edu.tr> with a value of 781.11, followed by Hacettepe University <hacettepe.edu.tr> (690.15), which is precisely the node which owns the maximum eigenvector (1) in the graph.

The dominance of universities in the betweenness centrality is notorious. In the 20 nodes with greater betweenness value, only four nodes correspond to companies (<zorlu.com.tr>; <pasabahce.com.tr>; <seas.gov.tr>; <brisa.com.tr>, although the eigenvector values of all of them are really very discrete (0.26, 0.08, 0.22, and 0.28 respectively).

As for the values of degree, the Middle East Technical University <metu.edu.tr> appears on the top position (94), followed by Ankara University <ankara.edu.tr> (90). Only 11 nodes get a degree value equal to 0 (all of them corresponding to companies). Full details of each node are accessible in Annex IV of the supplementary material 1.

As previously mentioned, the network is very sparse. Of the 15,750 possible combinations, only 7.4% (1,172) has obtained at least one hit. The combination of institutions most mentioned among each other corresponds to Suleyman Demirel University Turkey <sdu.edu.tr>, where are located 723,000 mentions towards Istanbul University <istanbul.edu.tr>. The Top 20 combinations with the highest number of hits (reflecting thus combination intensity) are shown in Table 7 (Top 100 is available in Annex V, supplementary material 1).

**Table 7. Top 20 combinations in the network according to URL mentions**

| URL MENTION | HITS | TYPE |
|---|---|---|
| FROM <sdu.edu.tr> TO <istanbul.edu.tr> | 723,000 | UNI |
| FROM <sdu.edu.tr> TO <atauni.edu.tr> | 152,000 | UNI |
| FROM <istanbul.edu.tr> TO <ankara.edu.tr> | 129,000 | UNI |
| FROM <sdu.edu.tr> TO <marmara.edu.tr> | 34,100 | UNI |
| FROM <istanbul.edu.tr> TO <hacettepe.edu.tr> | 33,100 | UNI |
| FROM <sdu.edu.tr> TO <ankara.edu.tr> | 27,600 | UNI |
| FROM <istanbul.edu.tr> TO <uludag.edu.tr> | 27,300 | UNI |
| FROM <sdu.edu.tr> TO <metu.edu.tr> | 24,800 | UNI |
| FROM <sdu.edu.tr> TO <sakarya.edu.tr> | 10,100 | UNI |
| FROM <sakarya.edu.tr> TO <istanbul.edu.tr> | 9,680 | UNI |
| FROM <sakarya.edu.tr> TO <ankara.edu.tr> | 6,820 | UNI |
| FROM <istanbul.edu.tr> TO <atauni.edu.tr> | 6,750 | UNI |
| FROM <sakarya.edu.tr> TO <sdu.edu.tr> | 6,110 | UNI |
| FROM <sakarya.edu.tr> TO <gazi.edu.tr> | 5,140 | UNI |
| FROM <sakarya.edu.tr> TO <selcuk.edu.tr> | 4,580 | UNI |
| FROM <istanbul.edu.tr> TO <marmara.edu.tr> | 3,960 | UNI |
| FROM <sakarya.edu.tr> TO <yildiz.edu.tr> | 3,870 | UNI |
| FROM <sdu.edu.tr> TO <ortaanadolu.com> | 3,540 | TRANSFER |
| FROM <sdu.edu.tr> TO <sabanciuniv.edu> | 3,040 | UNI |
| FROM <istanbul.edu.tr> TO <gazi.edu.tr> | 2,850 | UNI |

As shown in Table 7, the combinations with the highest number of hits correspond to entries from universities (type UNI). Of the 1,172 active combinations, 50.9% (597) correspond to mentions of this type, while 42.8% (502) correspond to combinations between a university and a company (type TRANSFER), and only 6.2% (73) to combinations between companies (type COM).

Despite the large number of combinations between universities and companies, the intensity of these relationships is very low (mean = 18.1) if compared to the UNI type relationships (2,155), although higher than those of COM type (mean = 10.4).

However, the symmetry of university-company relations is very unbalanced. Of the 502 combinations TRANSFER type, 90.6% (455) are URL mentions from university nodes to company nodes while only 9.4% (47) are directed from company nodes to universities. The most intense combination of these 47 combinations corresponds to the mentions directed from company node <aygaz.com.tr> to university node <ege.edu.tr>, with 71 mentions.

In Figure 2 we show the neighborhood of the node <etimaden.gov.tr>, which corresponds to the company associated with a greater number of universities (20 of 25), although the intensity of all these relations is very low.

**Figure 2. Neighbourhood of company node <etimaden.gov.tr> with university nodes**

Finally, Annex VI (supplementary material 1) shows by way of illustration a ranking of Turkish universities according to the degree of web interaction with Turkish companies. On the one hand we display the number of companies from which at least 1 mention is provided to or received from the corresponding university (labelled interaction degree), as well as the total number of mentions (interaction hit).

**4. Discussion**

All measures based on web indicators should be taken with some caution, in particular those obtained through the search commands of a general search engine such as Google, due to the high variability of the data and a number of inconsistencies, recently summarized by Willinson and Thelwall (2013).

Specifically, all measures based on hit count estimates (especially page count and URL mentions) should be treated cautiously since Google provides only rounded values. These limitations especially affect the use of these indicators to evaluate performance (e.g. the exact number of mentions to a URL). However, if used for relational purposes (e.g. to determine whether the relationship between "URL A" and "URL B" is greater than that between "URL A" and "URL Z", which is how it should be interpreted in this study), these limitations are minimised since all URLs are subject to the same error, and thus the error is statistically dispersed.

Nonetheless, the high figures in the URL mentions do pose a high error rate and a high variability over time. For this reason, two samples of URL mentions (November and December) were taken, finding no correlation between the two samples either for universities (r = 0.3) or companies (r = 0.4).

The regional search option (Turkey) logically provides much more discrete results and is therefore eligible for a lower error rate. The correlation between the data of November and December remains low - even better - for universities (r = 0.5), whereas it is positive and significant for companies (r = 0.9), a fact that also reflects a slowdown in the websites of companies.

In any case, certain errors and limitations of this regional search are observed. In the case of universities, based on the sample of December, three cases where the regional results are higher than the total results are detected: Dumlupinar University (total: 62,200; region: 685,000), Isik University (total: 43,000; region: 645,000) and Ordu University (total: 72,000; region: 475,000). In the November data, this effect was detected for 6 universities (only two of which coincide with those detected in December). For companies, data errors in November amounted to only two companies, however, increased to 11 in the December sample. It is likely that the separation of web visibility measures obtained through Open Site Explorer (domain authority, external inlinks and root domains) and URL mentions (See Table 4) may be due these inconsistencies.

We have also found specific differences (precisely for those URLs that present inconsistencies in the regional values shown above) between the HCE provided by <google.com> and that by <google.com.tr> (the one used in this work). For example, in the query <"bossa.com.tr" -site:bossa.com.tr>, 74,200 hits were obtained in the Turkish version of Google, but only 6,330 hits on Google (See Annex VII, supplementary material 1). This could be due to temporary mismatches in the timing of Google datacenters.

In any event, these limitations affect the general URL mentions more than the mentions specific between universities, which are orders of magnitude lower. Furthermore, the effect of these errors on the network is limited, not affecting the general overall appearance or the map results, though this data must be used with caution as they represent only and approximately the explicit relationships through their websites.

With regard to data about the Turkish universities' scholarly productivity (obtained from Scopus), we should also mention certain limitations that may have affected the data in some institutions:
- Most of the new universities formed after 2006, and the productivity is zero before that date. Concretely, from 2000 to 2004 we find 18 universities for which the number of citations received is "0", and for 8 universities, that number does not surpasses 10 citations.
- Scopus still has problems with the Turkish affiliations, which must be checked manually to be properly assigned. A similar limitation has been observed in a previous study which was conducted by Aytac (2010).
- The biased coverage of Scopus may affect on the performance of some Universities. For example, the medium of instruction in Galatasaray University is French, language less covered by Scopus (Orduna-Malea and Delgado López-Cózar 2014).

> Likewise, Mimar Sinan University is a Fine Art School, and Scopus is limited within that subject area.

Otherwise, correlations between measures about web visibility are higher in the university set than in the company set; these results are consistent with those previously obtained by Vaughan and Yang (2012), although largely depend on the different industrial sectors analysed. This confirms the need to compare homogeneous companies, as indicated by Vaughan and Wu (2004) and Vaughan, Gao and Kipp (2006). However, the number of firms in the sample by sector (10) is reduced, so that it would be expanded in the future to obtain more conclusive results.

Romero-Frias and Vaughan (2010) find that some financial variables (total assets, revenue and net income) shown to be correlated with external links since all of them are of cumulative nature. However, in this research, the product-based sales variable has not provided positive results broadly. By sector, its correlation with the web indicators has varied, highlighting the high values obtained for "Mining and quarrying" ($r = 0.77$) and "Metal products, machinery and equipment, professional instruments" ($r = 0.67$) sectors, unique in which the correlation was significant ($\alpha = 0.05$).

Otherwise, the companies measured by Cankir, Arslan and Seker (2015) with highest web reputation are mostly from banking sector while the lowest rates are from the mining sector. However, both the sample and the indicators used in that study are quite different (for example, any company from banking industry is included), therefore results cannot be compared.

The results should be interpreted not only at the technical level (accuracy, validity and reliability of indicators and sources), but conceptually. Obviously, the fact that a university and a company do not link each other on the Web does not prevent that they may maintain certain relationships or collaboration. In that case, the results of this study could help to solve this mismatch. It is also possible for universities and companies to maintain indirect relations through certain intermediaries such as science parks. Minguillo and Thelwall (2012), in their cybermetric analysis of the United Kingdom Science Park Association, show that the websites of science parks link to both universities and companies, serving as a bridge between them both in the offline and online levels. Special attention should be made with technoparks within large universities such as the Middle East Technical University (METU) and Bilkent University, which tend to maintain a greater level of interaction with companies, as the data of Annex VI show.

## 5. Conclusions

The indicators obtained (both for web presence and web visibility) indicate significant differences between the group of academic institutions and those related to companies within the web space of Turkey. This is particularly evident in the number of root domains (external websites from where at least one hyperlink to the studied domain is targeted). The set of universities receives, on average, links from 1,157 different websites while this indicator, in the case of companies, is reduced to only 81. This reflects a very low web impact of the Turkish industrial system on the Web regarding the university system. This result is considered of importance since the selected companies are the most distinguished in their respective sectors.

Moreover, we have obtained significant positive correlations ($\alpha = 0.01$) among all web indicators, both for all universities and companies. The following considerations are:
- In the case of universities, the page count indicators are more separated from web visibility indicators (citation data are also integrated in this category) than in the group of companies.
- The URL mentions are more correlated with page count indicators than with web visibility indicators, both in the case of universities and, more markedly in the companies. This is mainly due to the high correlation between the page count obtained in Google Scholar and URL mentions in universities ($r = 0.73$).
- In the case of companies, correlations heavily depend on sectors, obtaining very different values in each of the 10 sectors analysed. For example, the correlation between page count and Root domains is very high in some sectors (higher than $r = 0.5$ in 6 sectors) but very low in other sectors (lower than $r = 0.4$ in 4 sectors).

Finally, relations between the academic and business institutions obtained through URL mentions are very scarce. Of the 4,900 possible combinations between universities and companies (in both directions) to be obtained from the sample, only 502 (10.2%) obtains at least one hit (one mention between the two institutions). The intensity of relationships is very low as well, with an average of 18.1 hits per combination; in 90.6% of cases the mention occurs with the university as source and the company as target, and not the contrary. These results are reflected in the way the network perform, with companies on the periphery, surrounding the nucleus formed by universities, which have higher web presence and greater connectivity between them.

Given the limitations discussed in the previous section, we conclude that there is a web disconnection between the University and the industrial sector in Turkey. These data are especially relevant because in societies with high levels of innovation, interactions between universities and companies are usually elevated. In that sense, the results obtained in this study could serve as a starting point to determine the impact of the Turkish academic and industrial web systems and to identify the weakest interactions between these systems (and thus establishing possible strategic actions to strengthen them). Specifically, the interactions map via URL mentions could be used at the strategic level both as a barometer of the current interactions and as a roadmap to continuously enhance university-industry relations.

In any event, this research is exploratory and should be expanded in future studies with a larger sample of both universities and companies in each sector. Likewise, a longitudinal study rather than sectional would eliminate or smooth fluctuations of web data (especially URL mentions) as a more adequate understanding of the relations between Turkish institutions, and their web impact, is reached.

## 6. Notes
1. Scimago Institutions Ranking.
http://www.scimagoir.com (accessed 14 March 2015).

2. University - Industry Research Connections (UIRC)
http://www.cwts.nl/UIRC2014#Our_UIC_performance_indicators (accessed 14 March 2015).
3. Ranking web of World Universities.
http://www.webometrics.info/en/Europe/Turkey (accessed 14 March 2015).
4. URAP.
http://www.urapcenter.org/2014/country.php?ccode=TR (accessed 14 March 2015).
5. Supplementary material available at: FINAL URI TO INCLUDE IF ACCEPTED
6. Google Scholar.
http://scholar.google.com/intl/en/scholar/help.html#coverage (accessed 14 March 2015).
7. MOZ.
https://moz.com/researchtools/ose/ (accessed 14 March 2015).

## 7. References

Aguillo, I. F., Granadino, B., Ortega, J. L., & Prieto, J. A. (2006). Scientific research activity and communication measured with cybermetrics indicators. *Journal of the American Society for information science and technology*, *57*(10), 1296-1302.

Arslan, M.L., Seker, S.E. (2014). Web Based Reputation Index of Turkish Universities. *International Journal of E-Education E-Business E-Management and E-Learning*, *4*(3), 197-203.

Aytac, S. (2010). International scholarly collaboration in science, technology and medicine and social science of Turkish scientists. *The International Information & Library Review*, *42*(4), 227-241

Bahçıvan, E. (ed.) (2013). Turkey's Top 500 Industrial enterprises 2012. *The Journal of the Istanbul Chamber of Industry*, *48*(569), 1-124 [special issue].

Barabasi, A. L., & Albert, R. (1999). Emergence of Scaling in Random Networks. *Science*, *286*(5439), 509-512.

Cankir, B., Arslan, M. L., & Seker, S. E. (2015). *Web Reputation Index for XU030 Quote Companies. Journal of Industrial and Intelligent Information*, *3*(2), 110-113

Faba-Fernández, C., Guerrero-Bote, Vicente P., & Moya-Anegón, F. (2003). Data mining in a closed web environment. *Scientometrics*, *58*(3), 623-640.

Fruchterman, T. M., & Reingold, E. M. (1991). Graph drawing by force-directed placement. *Software: Practice and experience*, *21*(11), 1129-1164.

García-Santiago, L., & De Moya-Anegón, F. (2009). Using co-outlinks to mine heterogeneous networks. *Scientometrics*, *79*(3), 681-702.

Jolliffe, I. (2002). *Principal component analysis*. New York: Springer.

Khan, G. F., & Park, H. W. (2011). Measuring the triple helix on the web: Longitudinal trends in the university-industry-government relationship in Korea. *Journal of the American Society for Information Science and Technology*, *62*(12), 2443-2455.

Leydesdorff, L., & Etzkowitz, H. (1996). Emergence of a Triple Helix of university—industry—government relations. *Science and public policy*, *23*(5), 279-286.

Leydesdorff, L., & Park, H. W. (2014). Can Synergy in Triple-Helix Relations be Quantified? A Review of the Development of the Triple-Helix Indicator. *arXiv preprint arXiv:1401.2342*.

Meyer, M. (2000). What is special about patent citations? Differences between scientific and patent citations. *Scientometrics*, *49*(1), 93-123.

Meyer, M., Siniläinen, T., & Utecht, J. T. (2003). Towards hybrid Triple Helix indicators: A study of university-related patents and a survey of academic inventors. *Scientometrics*, *58*(2), 321-350.

Minguillo, D., & Thelwall, M. (2012). Mapping the network structure of science parks: An exploratory study of cross-sectoral interactions reflected on the web. *Aslib Proceedings*, *64*(4), 332-357.

Montesinos, P., Carot, J.M., Martínez, J.M. & Mora, F. (2008). Third Mission Ranking for World Class Universities: beyond teaching and research. *Higher Education in Europe*, *33*(2/3), 259-271.

Orduna-Malea, E., & López-Cózar, E. D. (2014). Google Scholar Metrics evolution: an analysis according to languages. *Scientometrics*, *98*(3), 2353-2367.

Ortega, J. L., Orduna-Malea, E., & Aguillo, I. F. (2014). Are web mentions accurate substitutes for inlinks for Spanish universities?. *Online Information Review*, *38*(1), 59-77.

Priem, J., & Hemminger, B. H. (2010). Scientometrics 2.0: New metrics of scholarly impact on the social Web. *First Monday*, *15*(7). Available at http://firstmonday.org/ojs/index.php/fm/article/viewArticle/2874 (accessed 31 December 2014).

Romero-Frías, E. (2011). Googling Companies-a Webometric Approach to Business Studies. *Leading Issues in Business Research Methods*, *7*(1), 93-106.

Stuart, D., & Thelwall, M. (2006). Investigating triple helix relationships using URL citations: a case study of the UK West Midlands automobile industry. *Research Evaluation*, *15*(2), 97-106.

Thelwall, M. (2004). *Link analysis: an information science approach.* San Diego: Academic Press.

Thelwall, M. (2014). A brief history of altmetrics. *Research trends*, *37*, 3-4. Available at http://www.researchtrends.com/issue-37-june-2014/a-brief-history-of-altmetrics/ (accessed 31 December 2014).

Thelwall, M., & Harries, G. (2003). The connection between the research of a university and counts of links to its Web pages: An investigation based upon a classification of the relationships of pages to the research of the host university. *Journal of the American Society for Information Science and Technology*, *54*(7), 594-602.

Vaughan, L. (2004). Exploring website features for business information. *Scientometrics*, *61*(3), 467-477.

Vaughan, L. (2006). Visualizing linguistic and cultural differences using Web co-link data. *Journal of the American Society for Information Science and Technology*, *57*(9), 1178-1193.

Vaughan, L. and Wu, G. (2004). Links to commercial web sites as a source of business information. *Scientometrics*, *60*(3), 487-96.

Vaughan, L., & Romero-Frías, E. (2012). Exploring web keyword analysis as an alternative to link analysis: a multi-industry case. *Scientometrics*, *93*(1), 217-232.

Vaughan, L., & Thelwall, M. (2003). Scholarly use of the web: what are the key inducers of links to journal web sites?. *Journal of the American Society for Information Science and Technology*, *54*(1), 29-38.

Vaughan, L., Gao, Y. and Kipp, M. (2006). Why are hyperlinks to business Websites created? A content analysis. *Scientometrics*, *67*(2), 291-300.

Wilkinson, D., & Thelwall, M. (2013). Search markets and search results: The case of Bing. *Library & Information Science Research*, *35*(4), 318-325.

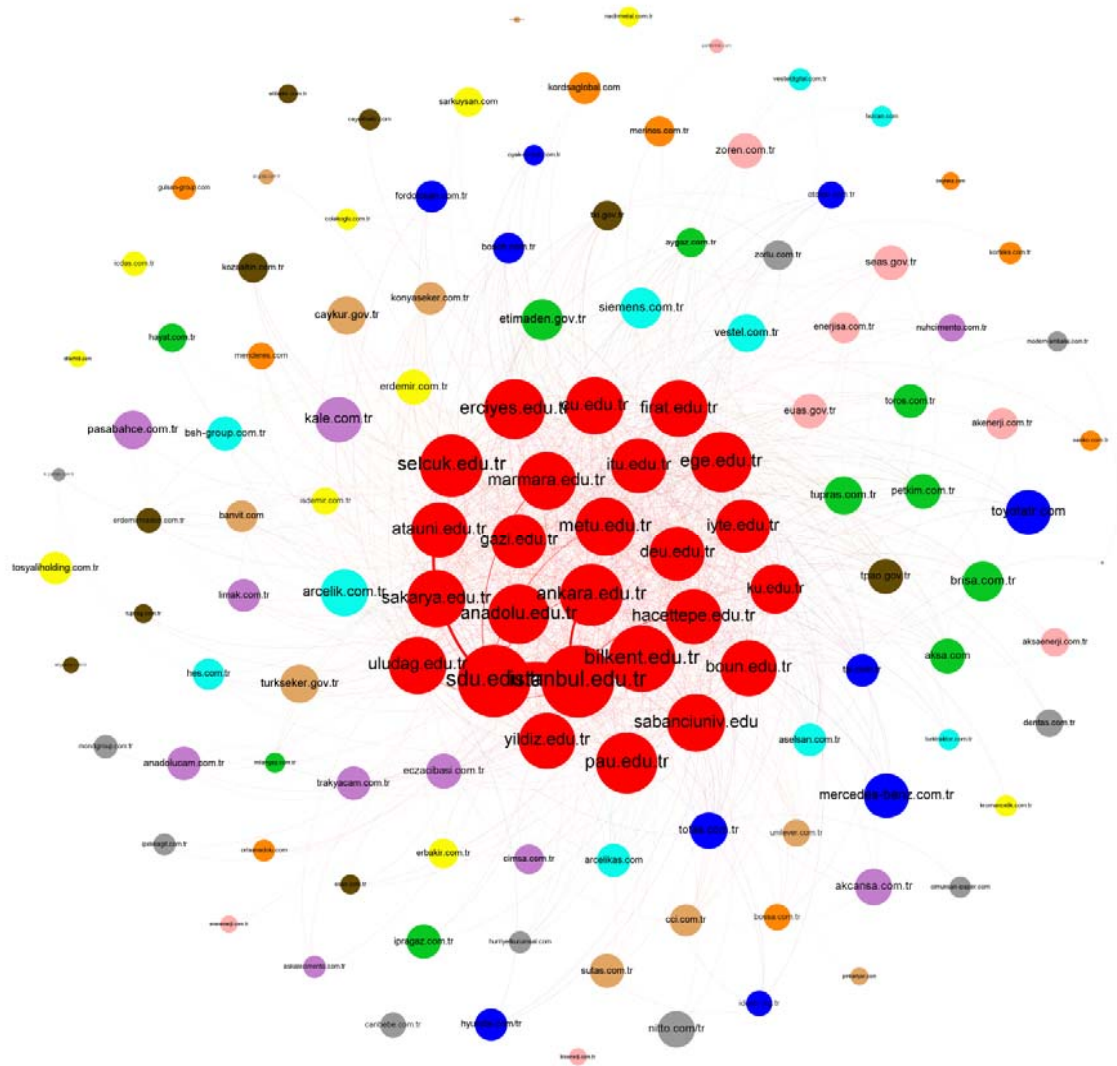

**Figure 1. University and company interaction network (n=123; algorithm: Fruthterman-Reingold)**

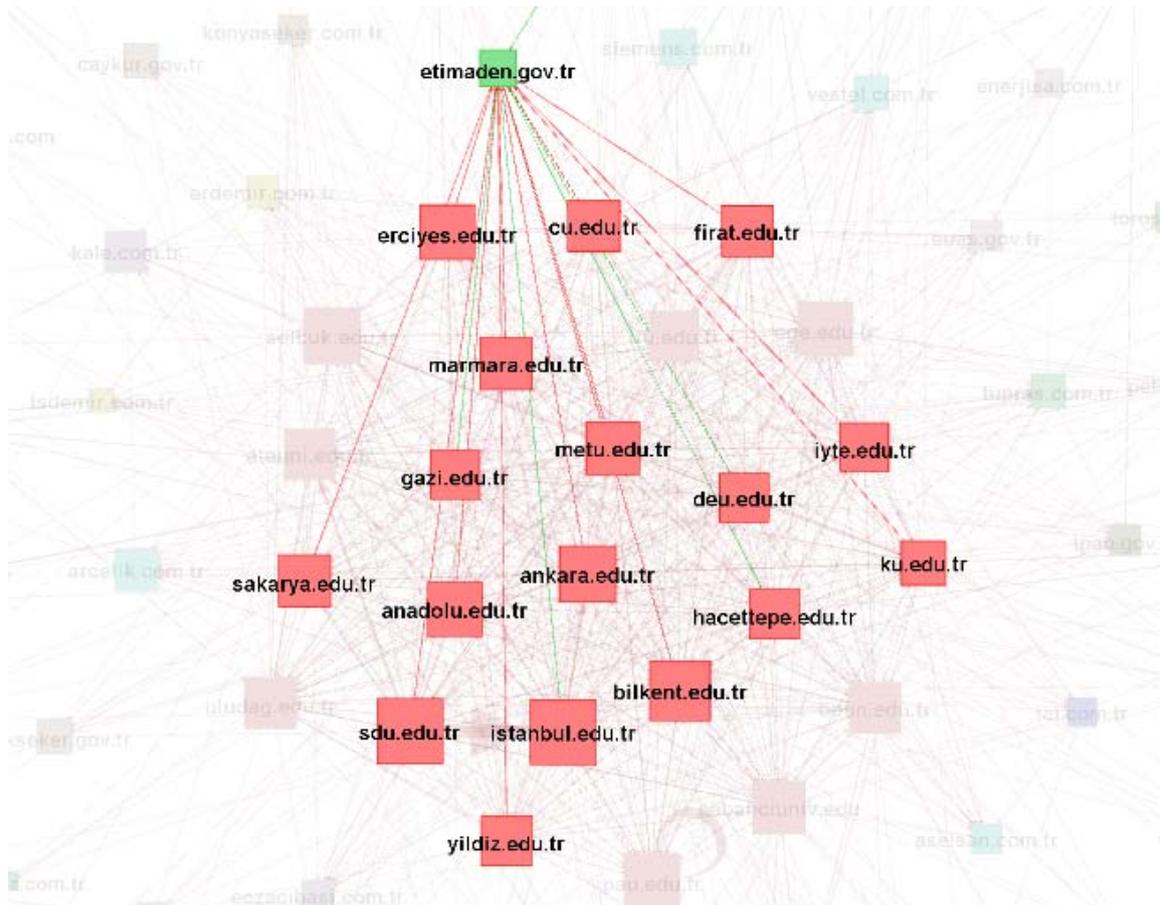

**Figure 2. Neighbourhood of company node <etimaden.gov.tr> with university nodes**